\documentclass[sigconf, screen]{acmart}

\AtBeginDocument{%
  }

\copyrightyear{2026}
\acmYear{2026}
\setcopyright{cc}
\setcctype{by-nc-sa}

\acmConference[Next Steps for Augmented Reality On-the-Move: Challenges \& Opportunities at CHI '26]{Next Steps for Augmented Reality On-the-Move: Challenges \& Opportunities @ CHI '26}{April 16, 2026}{Barcelona, Spain}
\acmBooktitle{Next Steps for Augmented Reality On-the-Move: Challenges \& Opportunities at CHI '26}

\acmDOI{}                          % blank ⇒ nothing is printed

\acmISBN{}                         % empty string  → ISBN disappears
\acmPrice{}                        % optional: removes the "$15.00" line

\usepackage{enumitem}

\begin{document}

\title[Toward Governing Perception in MR on the Move]{Toward Governing Perception in Safety-Critical Mediated Reality on the Move}

\thanks{Position paper presented at Next Steps for Augmented Reality On-the-Move: Challenges \& Opportunities - Workshop at CHI~’26}

\author{Pascal Jansen}
\email{pascal.jansen@uni-ulm.de}
\orcid{0000-0002-9335-5462}
\affiliation{%
  \institution{Institute of Media Informatics, Ulm University}
  \city{Ulm}
  \country{Germany}
}

\renewcommand{\shortauthors}{Pascal Jansen}

\begin{abstract}
Wearable Augmented Reality (AR) is increasingly deployed in on-the-move contexts such as automated driving, cycling, and pedestrian navigation. To date, most systems rely on additive overlays that highlight hazards, intentions, or predictions without altering the scene itself. However, advances in head-mounted displays and computer vision now enable Diminished and Modified Reality techniques that suppress, transform, or substitute scene elements. These capabilities conceptually extend AR into Mediated Reality (MR), shifting the design space from “what to add” to “what is perceptually available.” Because such mediation reshapes the evidential basis for situation awareness and trust calibration, it raises novel interaction challenges. This position paper argues that MR on the move must become governable, as users need mechanisms to configure, inspect, and understand mediation without compromising safety. Additionally, this position paper outlines design challenges related to governance granularity, epistemic signaling, and accountability, and frames MR on the move as a research agenda for governable perceptual mediation in dynamic, safety-critical environments.
\end{abstract}

\begin{CCSXML}
<ccs2012>
 <concept>
  <concept_id>10003120.10003121.10003124.10010865</concept_id>
  <concept_desc>Human-centered computing~Human computer interaction (HCI)</concept_desc>
  <concept_significance>500</concept_significance>
 </concept>
 <concept>
  <concept_id>10003120.10003121.10011748</concept_id>
  <concept_desc>Human-centered computing~Interaction paradigms</concept_desc>
  <concept_significance>300</concept_significance>
 </concept>
 <concept>
  <concept_id>10003120.10003121.10003124.10010866</concept_id>
  <concept_desc>Human-centered computing~Ubiquitous and mobile computing</concept_desc>
  <concept_significance>200</concept_significance>
 </concept>
 <concept>
  <concept_id>10003120.10003121.10003124.10010868</concept_id>
  <concept_desc>Human-centered computing~Mixed / augmented reality</concept_desc>
  <concept_significance>200</concept_significance>
 </concept>
</ccs2012>
\end{CCSXML}

\ccsdesc[500]{Human-centered computing~Human computer interaction (HCI)}
\ccsdesc[300]{Human-centered computing~Interaction paradigms}
\ccsdesc[200]{Human-centered computing~Ubiquitous and mobile computing}
\ccsdesc[200]{Human-centered computing~Mixed / augmented reality}

\keywords{augmented reality, mediated reality, situation awareness, trust calibration, mobility}

\maketitle

\section{Introduction}

Wearable Augmented Reality (AR) systems are increasingly used in on-the-move contexts such as pedestrian navigation, cycling, and automated vehicles \cite{stefanidi2024augmented}. In these contexts, users move through environments that change continuously \cite{jansen2023autovis}. For example, vehicles approach at different speeds, pedestrians hesitate before crossing, cyclists appear from behind obstacles, and traffic density varies. Under such conditions, users must quickly judge what is present, what others intend to do, and what will likely happen next. These judgments directly affect safety \cite{endsley1995sa}.

To support such decisions, much automotive AR research has focused on additive overlays. These overlays highlight detected objects, visualize predicted trajectories, or display maneuver intentions to increase transparency (e.g., \cite{colley2021effects, jansen2024visualizing, jansen2025longitudinal}). Similarly, AR systems for vulnerable road users mainly rely on additive visual signals layered onto the environment without altering it \cite{tabone2023augmented, stefanidi2024augmented}.

However, recent advances in computer vision models (e.g., object detection and inpainting \cite{kari2021transformr}) have extended the capabilities of Head-Mounted Displays (HMDs) beyond AR. Following the definition of Mediated Reality (MR) by \citet{mann2002mediated}, HMD systems can now additionally suppress or remove parts of the scene (Diminished Reality, DR) or transform and substitute them (Modified Reality, ModR) \cite{lindlbauer2018remixed, cheng2022towards}. The potential applications are manifold. For example, roadside clutter (e.g., advertisements) could be removed, intersections could be visually simplified (e.g., by removing non-relevant vehicles for drivers or by simplifying visually cluttered crossings for pedestrians), or predicted trajectories could be emphasized for specific road users (e.g., pedestrians at road crossings).
\citet{jansen2026mirage} recently demonstrated that AR, DR, and ModR can be integrated as Automotive Mediated Reality (AMR) and deployed in real traffic within a moving vehicle. Expert evaluations indicated that AMR effects were interpreted not merely as visual refinements, but as interventions that influenced judgments of environmental complexity, risk, and system capability \cite{jansen2026mirage}.

These observations suggest that DR and ModR are not only technical extensions, but interaction-level variables that reshape how users interpret the scene. While AR primarily influences saliency, DR and ModR influence perceptual availability, which is the evidential basis from which users construct SA \cite{endsley1995sa}. SA unfolds in three stages: perceiving relevant elements (Level 1), comprehending their meaning (Level 2), and projecting future states (Level 3). For example, if a DR application removes an element, it never enters SA Level 1. If predicted trajectories are emphasized without distinction from sensor observations, users may interpret forecasts as certain rather than probabilistic.

As mobility contexts involve automated decision-making, these shifts in SA extend to users’ appraisals of the traffic scene and road users' capabilities (e.g., automated vehicles). For instance, to form trust in automation, users evaluate whether the vehicle reliably detects relevant actors, accurately predicts their motion, and plans appropriate responses \cite{LeeSee2004Trust}. Trust calibration depends on aligning reliance with actual capability under uncertainty \cite{LeeSee2004Trust}. Consequently, when MR interfaces mediate environmental complexity (e.g., removing road users), users may misjudge what the automation can detect, forecast, or handle, leading to non-calibrated trust \cite{jansen2025longitudinal}.

This position paper, therefore, argues that AR in on-the-move contexts will evolve into MR interfaces, and must incorporate mechanisms that make perceptual mediation \textit{governable}. Rather than treating scene element suppression and transformation as hidden rendering choices, interfaces can expose the epistemic status of scene elements, for example, by differentiating directly sensed road users, algorithmic interpretations, predicted trajectories, and intentionally filtered content through visual encodings or interactive inspection techniques. Such mechanisms enable users not only to see what is emphasized, altered, or removed, but also to understand how the scene has been constructed and, where appropriate, adjust mediation levels towards forming appropriate SA and appraisal. 
Insights from this setting may inform the broader design of MR systems in other dynamic, safety-relevant mobility domains, such as pedestrian navigation and micromobility.

\section{Design Challenges for Governable Mediated Reality on the Move}

Governance of MR in on-the-move contexts concerns interaction design choices that define (i) when users can influence mediation, (ii) what they can control, (iii) how epistemic status is communicated, and (iv) how responsibility is attributed when mediation changes what was perceivable in the first place.

\subsection{When and How Can Users Govern Mediation?}

Continuous in-situ governance of MR during driving, walking, or cycling is typically infeasible because interaction is mechanically constrained, and additional controls risk increasing workload or diverting attention from primary safety tasks. This contrasts with AR systems in stationary rooms \cite{lindlbauer2018remixed}, where exploratory inspection is often feasible.

An interaction model with three loci of control might address when and how governance can occur. First, \textbf{pre-ride configuration}: users select high-level mediation profiles (e.g., “reduced clutter,” “highlight vulnerable road users,” “conservative uncertainty display”) via low-demand interfaces such as a companion app, a dashboard menu, or a voice-based setup dialogue before entering motion. Interaction modalities may include touch- or gesture-based selection and conversational voice interaction (e.g., “prioritize pedestrians today”). Second, \textbf{minimal in-situ override}: during motion, only lightweight interventions are permitted, such as a single voice command (“show full scene”), a gaze-triggered reveal of a suppressed object, or a context-attached affordance directly anchored to the mediated element (e.g., a subtle icon next to an inpainted vehicle that expands on fixation). These mechanisms might constrain interaction cost while preserving agency. Third, \textbf{post-ride inspection}: users review a structured mediation summary that explains when and why transformations occurred.

This temporal separation aligns with prior work on inspectability and reviewability in adaptive systems \cite{xu2023xair}, yet mobility particularly introduces a binding problem: users must connect a post-hoc explanation to a moment experienced while in motion and under limited attention. To support this, mediation events could be logged and exposed through event-based replay mechanisms that bind each intervention to spatiotemporal anchors (timestamp, location, maneuver context; see \cite{jansen2023autovis}). The interface could provide synchronized dual views (mediated vs. unmediated, or delta visualization), annotated with triggers and system confidence levels. Hence, (i) at what moments (e.g., low traffic density, low speed, straight-road segments) is it safe and appropriate to surface even lightweight mediation controls without degrading performance, and (ii) which replay and annotation formats best support recognition and understanding of past mediation events that were originally experienced under attentional load?

\subsection{Granularity of Mediation Governance}

To govern AR and visualization systems, users often select between presets (e.g., overview vs. detail mode) or adjust individual visualization parameters \cite{zollmann2020visualization}. In mobility, however, granularity is coupled with safety because users cannot meaningfully “debug” the world while moving. Moreover, non-expert users may not understand the MR system's internal models or training data \cite{jansen2024visualizing}. Consequently, overly fine control risks misconfiguration, while overly coarse control risks epistemic opacity.

\textbf{Coarse control} (e.g., “simplified environment” vs. “full detail”) could be used. For instance, a “simplified environment” mode might suppress parked vehicles, reduce texture and facade detail, and collapse distant road users into abstract glyphs, whereas “full detail” would render the scene without suppression or abstraction. These modes might be easier to communicate, but they can obscure which aspects are being altered, such as object density (number of rendered agents) or semantic emphasis (which object classes are highlighted). Users may therefore not know whether, for example, background clutter reduction can also affect peripheral hazard visibility.

\textbf{Fine-grained control} (e.g., “suppress background clutter only,” “de-emphasize predicted trajectories,” “highlight vulnerable road users”) may increase agency and allow alignment with individual risk preferences, but it risks complexity and misconfiguration \cite{jansen2025opticarvis}. A mixed-initiative approach could allow the system to propose adjustments based on context (e.g., high traffic density or reduced visibility), with the user confirming or overriding the proposal. However, this raises governance demands, as proposals should be intelligible (e.g., “why is clutter suppression suggested now?”), reversible, and attributable to concrete triggers (e.g., “activated due to high pedestrian density”) to achieve user acceptance \cite{liao2020questioning, poursabzi2021manipulating}.

This framing yields two research questions: (i) how does control granularity affect SA and appraisal (e.g., trust, perceived safety) when mediation changes perceptual availability under time pressure? (ii) Which UI mechanisms, such as delta visualizations, explanations, or context-attached annotations, make “what exactly changed” sufficiently legible that coarse controls do not become epistemically opaque?

\subsection{Epistemic Signaling of Mediated Reality in On-the-Move Contexts}

Signaling the epistemic status of MR must be compatible with motion constraints. Persistent legends or dense encodings may clutter the scene and compete with hazard perception. Prior work on uncertainty visualization and explainable interfaces has proposed various techniques, such as persistent vs. on-demand views and progressive disclosure (e.g., see \cite{jansen2024visualizing, kay2016ish}).

These point to interaction techniques that minimize sustained visual load, such as gaze-triggered inspection and transient overlays on demand. However, these techniques may fail in some scenarios. For example, on-demand inspection assumes users detect uncertainty early enough to request clarification, which may not hold in fast-moving scenes, while persistent encodings can produce visual noise and inattentional blindness \cite{wickens2008multiple}. This motivates the question: how can epistemic signaling remain legible and behaviorally effective (i.e., actually change interpretation and action) without degrading scene clarity in dynamic, safety-critical environments?

\subsection{Responsibility and Attribution of Mediated Reality in Automated Mobility}

In automated mobility, MR that suppresses or transforms environmental cues changes the SA available to human intervention, thereby directly affecting how responsibility is attributed between the user and the automation. Prior work primarily focuses on communicating internal model reasoning and decision processes, typically assuming a stable perceptual representation of the environment \cite{jansen2024visualizing, jansen2025longitudinal, ehsan2021expanding}. Governable MR adds a distinct layer because the interface can visually alter the world itself. This is critical in mobility because small perceptual changes might alter users' hazard detection and emergency response times.

MR systems could log suppression/transformation events, associated triggers, confidence levels, and user-facing signals displayed at the time. Users (and potentially third parties) should be able to retrospectively inspect what was mediated during an incident, ideally via a delta that clarifies what was removed, modified, or emphasized. This yields a research question: what logging and replay abstractions provide actionable accountability without overwhelming users, while remaining faithful enough to support post-incident interpretation of whether sufficient perceptual access existed?

\section{Conclusion}

This position paper argues that increasing advances in HMD technology and computer vision models will conceptually extend AR in on-the-move contexts into MR through DR and ModR techniques (e.g., see AMR \cite{jansen2026mirage}). While AR overlays remain effective for highlighting hazards and intent, adding information can increase clutter and compete for limited user attention. Hence, MR complements AR by enabling selective suppression, transformation, or abstraction of elements in safety-critical contexts. This could enable semantically richer applications that can improve users’ SA and, thereby, positively influence traffic safety and user appraisal (e.g., trust calibration) in on-the-move contexts such as automated mobility, pedestrian navigation, and micromobility.

However, MR introduces challenges for interaction, as users no longer interact with the world directly but with an MR system-constructed representation. Governability, intelligibility, and auditability, therefore, can become central design requirements toward MR technology acceptance, as users must be able to understand what was altered, why it was altered, and how to influence these processes.
This reframing positions MR in on-the-move contexts as a broader research agenda on governable perceptual mediation in dynamic environments. 
%Moreover, addressing these interaction challenges becomes increasingly relevant as mobility systems become more automated.

%\begin{acks}
% To Robert, for the bagels and explaining CMYK and color spaces.
%\end{acks}

\bibliographystyle{ACM-Reference-Format}
\bibliography{mediated-reality-on-the-move}

% \appendix

\end{document}